# Compositional Control and Optimization of Molecular Beam Epitaxial Growth of $(Sb_2Te_3)_x(MnSb_2Te_4)_y$ Magnetic Topological Insulators

*Ido Levy,[1,2] Candice Forrester,[1,2] Haiming Deng,[3,4] Manuel Goldan,[5] Martha R. McCartney,[5] David J. Smith,[5] Christophe Testelin,[6] Lia Krusin-Elbaum,[3,4] and Maria C. Tamargo\*,[1,2]*

[1]Department of Chemistry, The City College of New York, NY

[2]Ph.D. Program in Chemistry, The Graduate Center of the City University of New York, NY

[3]Department of Physics, The City College of New York, NY

[4]Ph.D. Program in Physics, The Graduate Center of the City University of New York, NY

[5]Department of Physics, Arizona State University, Tempe, AZ

[6]Sorbonne Université, CNRS, Institut des NanoSciences de Paris, Paris, France

Abstract

Magnetic topological insulators such as $MnBi_2Te_4$ and $MnSb_2Te_4$ are promising hosts of novel physical phenomena such as quantum anomalous Hall effect and intrinsic axion insulator state, both potentially important for the implementation in topological spintronics and error-free quantum computing. In the bulk, the materials are antiferromagnetic but appropriate stacking with non-magnetic layers or excess Mn in the crystal lattice can induce a net ferromagnetic alignment. In this work we report the growth of $(Sb_2Te_3)_x(MnSb_2Te_4)_y$ layers with varying Mn content by molecular beam epitaxy. The Mn flux fraction provided during growth controls the percent of $MnSb_2Te_4$ that is formed in the resulting layers by a self-assembly process. Highly crystalline layers with compositions varying between $Sb_2Te_3$ (y=0) and $MnSb_2Te_4$ (x=0) were obtained. The results show that Mn incorporates as a structural component to form $MnSb_2Te_4$, and as an impurity element both in $Sb_2Te_3$ and in $MnSb_2Te_4$. Two modifications of the growth conditions were implemented to enhance the incorporation of Mn as a structural element to form $MnSb_2Te_4$. Annealing of a thin portion of the layer at the beginning of growth (pre-anneal step), and increasing the growth temperature, both result in a larger percent of $MnSb_2Te_4$ for similar Mn flux fractions during growth. Samples having at least a few percent of $MnSb_2Te_4$ layers exhibit ferromagnetic behavior likely due to the excess Mn in the system which stabilizes on Sb sites as $Mn_{Sb}$ antisite defects.

Introduction:

Incorporation of magnetic elements into 3D topological insulators (TIs) such as $Bi_2Te_3$ and $Sb_2Te_3$ has garnered a great deal of interest in recent years due to the potential to achieve quantum anomalous Hall effect (QAHE)[1]. QAHE was first demonstrated by incorporating Cr as a dopant in thin (Bi,Sb)$_2$Te$_3$ films[2]. However, defects caused by the incorporation of the Cr atoms into the TI crystal lattice resulted in the effect being only evident at sub-Kelvin temperatures[3-6]. Recently it was discovered that crystal growth of TIs, specifically Bi$_2$Te$_3$ and Sb$_2$Te$_3$, with Mn leads to the formation of a new crystal phase, MnBi$_2$Te$_4$ and MnSb$_2$Te$_4$, respectively, having a septuple layer (SL) structure rather than the typical quintuple layer (QL) structure of the non-magnetic TI[7]. The resulting materials are intrinsically magnetic rather than magnetically doped and may lead to fewer defects. A single SL is ferromagnetic, with the magnetic spins aligned out of plane. However, when stacked to form a bulk structure, the SLs couple antiferromagnetically and are not conducive to the observation of QAHE, except in the ultrathin layers of odd number (5-7) of SLs[8]. One approach used to resolve this problem was to separate the magnetic SLs layers with a few non-magnetic QLs, enabling ferromagnetic coupling between the non-adjacent SLs. The QAHE was recently demonstrated in a 1:4 MnBi$_2$Te$_4$:Bi$_2$Te$_4$ system at several Kelvin[9]. Another approach recently shown to produce ferromagnetic coupling of the SLs in MnSb$_2$Te$_4$ to introduce excess Mn into the pure SL structure. It has been proposed that Mn$_{Sb}$ antisite defects induce ferromagnetic coupling between SLs[10-11].

The growth of MnSb$_2$Te$_4$, or of mixed SL:QL layered structures, has been reported for bulk growth techniques as well as epitaxial growth techniques[12-14]. In most cases, the formation of SLs occurs by self-assembly depending on the amount of Mn provided during growth. Typically, to grow these materials, all the elements are mixed simultaneously in order to form the crystal, and the resulting mix of SLs and QLs depends on the amount of Mn added. Although this is a convenient method to grow these materials due to its simplicity, it has some important drawbacks, particularly the difficulty to control the formation of a specific structure or composition reproducibly and consistently throughout the crystal lattice. Molecular beam epitaxy (MBE) with its layer-by-layer growth mechanism promises to provide certain benefits for control of the desired structures. However, little work has yet been done to systematically investigate and understand the MBE growth parameters that affect the structural properties of these materials. Such an understanding is essential for the intentional growth of precise layered structures with accurate compositional control and the desired magnetic properties.

In this work we report the growth of $(Sb_2Te_3)_x(MnSb_2Te_4)_y$ structures by MBE. Variation of the Mn flux during growth enables control of the SL to QL ratio in the layer, although competing processes, such as Mn dopant incorporation in the QL and excess Mn in the SLs result in irreproducible and less than accurate control. Moreover, the distribution of the SLs throughout the crystal is random so that the local properties of the material vary. Modified growth techniques, involving the implementation of an annealing step in the early stages of growth, and increasing the growth temperature are explored and the results show improved crystal quality and a preferential incorporation of Mn into SLs resulting from these changes. A mechanism is proposed to explain and understand the observed improvements in the control of growth and the materials properties.

Experimental:

All samples were grown in a Riber 2300P system with base pressure of $3-5 \times 10^{-10}$ Torr, equipped with reflection high-energy electron diffraction (RHEED) for *in-situ* growth monitoring, on epi-ready c-plane (0001) sapphire substrates. The substrates were heated in vacuum to 600°C for 1 hour prior to growth to remove impurities from the surface. High purity 6N Antimony (Sb), Tellurium (Te) and 5N8 Manganese (Mn) fluxes were achieved by a Riber double zone cell for Sb and single zone Knudsen cells for Mn and Te. The fluxes were measured by the beam equivalent pressure (BEP) read by a Riber ion gauge placed in the position of the substrate prior to growth. The Mn BEP fraction, BEP(Mn)/[BEP(Mn) + BEP(Sb)], proportional to the Mn flux fraction, was used to control the Mn content during growth, and was varied between 0.00 and 0.11. Growth was performed under excess Te flux. The samples were grown by MBE via a 2-step growth method consisting of an initial deposition of a thin (usually 3-5 nm) low temperature buffer (LTB) $Sb_2Te_3$ layer, at 200°C for ~3-5 minutes. The LTB layer growth was then stopped, while keeping constant Te flux at the sample surface, and the substrate temperature was raised to the desired growth temperature of 245 - 255°C. A LTB growth has been shown to improve nucleation in MBE growth of TIs[15-16]. Next, the Mn containing layer was grown for 1-2 hours, during which the Mn, Sb and Te sources were simultaneously impingent on the sample surface. The Te:Sb BEP ratios were kept between 20-30, ensuring excess Te during growth. All samples used similar Sb flux, and the composition was varied by adjusting the Mn cell temperature to change the Mn flux. The growth rates for the samples varied between 0.4 and 1.2 nm/min depending on the amount of Mn incorporated.

All sample were characterized by a variety of post growth techniques. High resolution X-Ray diffraction (HR-XRD) measurements were performed using a Bruker D8 Discover diffractometer with a da-Vinci configuration and a Cu Kα1(1.5418 Å) source. Atomic force microscopy (AFM) images were obtained using Bruker Dimension FastScan AFM with a FastScan-A silicon probe. Scanning transmission electron microscope (STEM) images (Figures 2b-e) were performed (EAG Laboratories) using a Hitachi HD-2700 Spherical Aberration Corrected Scanning-TEM system coupled with energy dispersive X-ray spectroscopy (EDX). Further high-resolution

STEM imaging (Figures 2a and 6) was carried out using a probe-corrected JEOL ARM-200F operated at 200 keV. Hall transport measurements were performed in a 14 T Quantum Design physical property measurement system (PPMS) in 1 mTorr (at low temperature) of He gas. Electrical contacts in the vdP configuration were made with indium bonded on the edge of the thin film.

Results and Discussion:

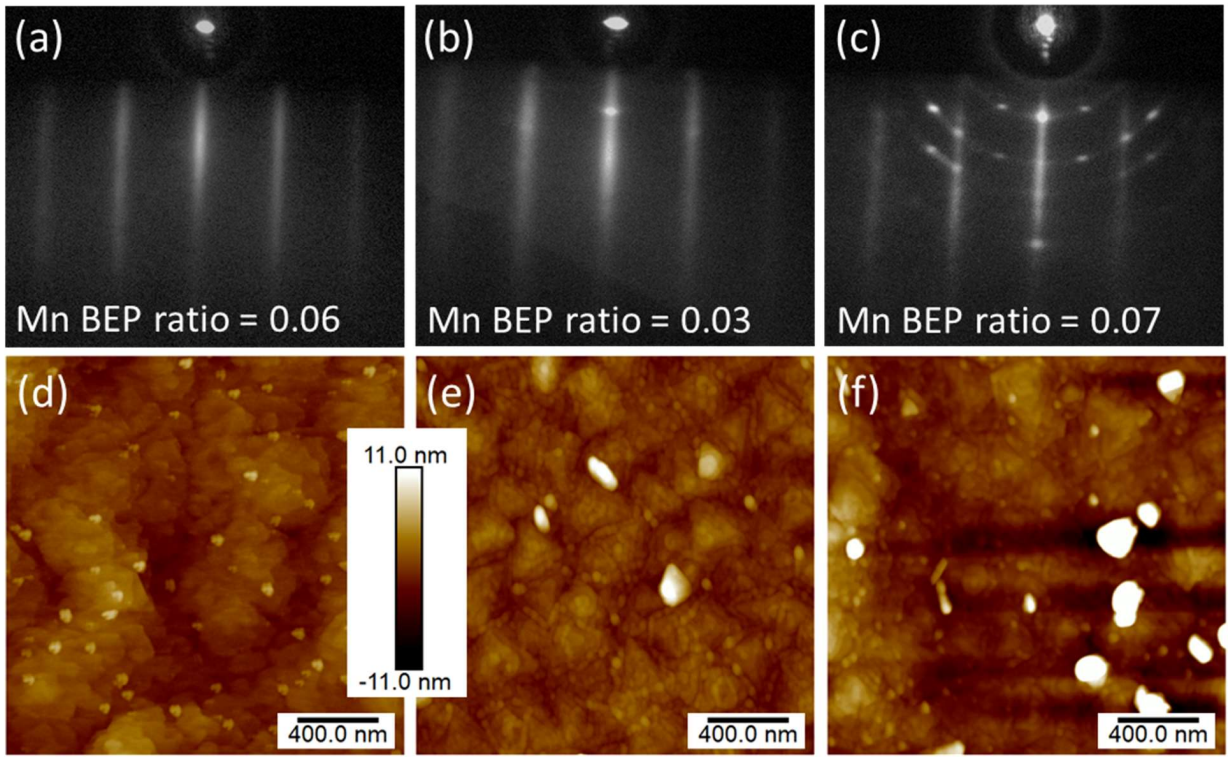

Figure 1: RHEED images (a-c) of three samples and their corresponding 2μm x 2μm AFM (d-f) images. The Mn BEP ratios for each sample are noted in the RHEED image. All AFM images are based on the same scale bar noted between (d) and (e).

Growth was monitored *in-situ* by reflection high energy electron diffraction (RHEED). Typical patterns at the end of the growth, and atomic force microscopy (AFM) surface images are shown in Figure 1. The initial RHEED pattern after LTB and during initiation of growth often exhibited some polycrystalline rings or spotty features which typically disappeared or became less prominent as the growth proceeded. Figures 1a-c show the RHEED images for three samples grown with different Mn BEP ratios, taken at the end of the growth. Most samples, regardless of composition, showed a streaky pattern, such as the one shown in Figure 1a; however, a few samples exhibited additional features suggesting rough surfaces and some disorder, as illustrated in Figures 1b-c. A corresponding variation in surface roughness is also visible in the AFM images presented in Figures 1d-f.

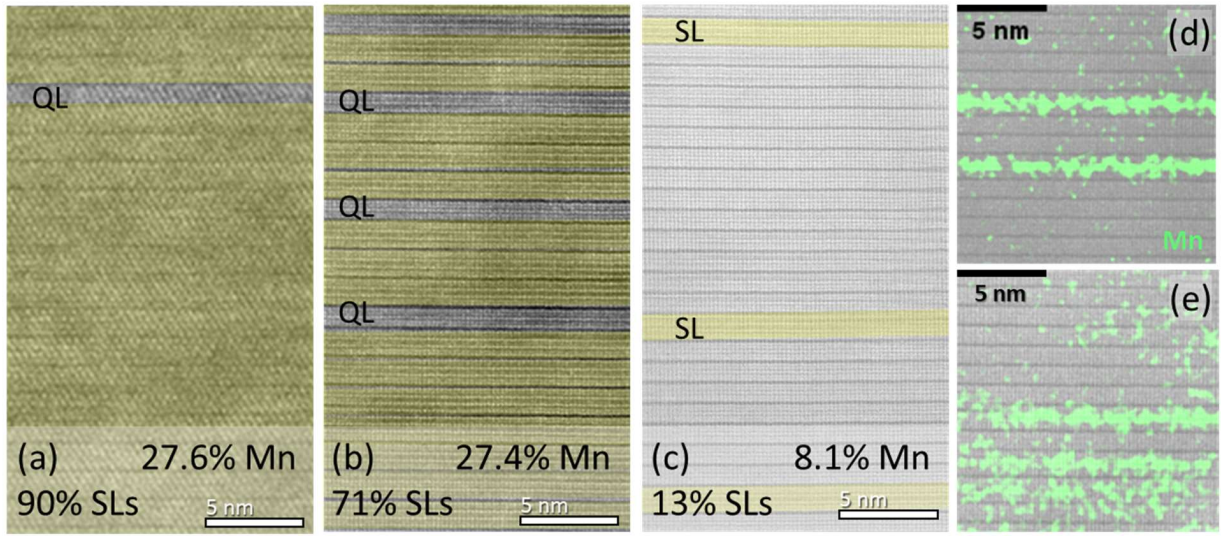

Figure 2: HAADF STEM images of three samples grown with different Mn BEP ratios, 0.11, 0.10 and 0.02 for (a), (b) and (c) respectively. Septuple layers (SLs) are highlighted with yellow and quintuple layers (QLs) are in grey. (a) Sample containing 27.6% Mn (obtained from EDX measurements) and 90% SLs, (b) Sample containing 27.4% Mn and 71% SLs, (c) a sample

containing 8.1% Mn and 13% SLs. (d,e) Cross sectional EDX scan of Mn (green) overlaid on two regions of the TEM image from the sample shown in (c).

Scanning transmission electron microscopy (STEM) observations were performed on some of the samples, primarily using high-angle annular-dark-field (HAADF) imaging because of its high-Z sensitivity. Figure 2 compares HAADF STEM cross sections of three samples having different Mn levels. The low magnification STEM images presented in Figures 2a-c enabled the analysis of large regions of each sample and estimates for the relative amount of septuple layers (%SL). The results illustrate the formation of SLs and QLs in different proportions, ranging from 90% SLs in Figure 2a to 13% SLs in Figure 2c. All of the STEM images show well ordered, highly crystalline structures and no obvious defects. The sample of Figure 2a, which was grown using a Mn BEP fraction of 0.11, has a crystal structure that consists almost completely of SLs, 90% calculated from the STEM image, and has a 27.6% Mn atomic percent, as obtained from energy dispersive X-ray (EDX). This result implies that, in addition to the Mn incorporated to form the SL structure, a significant amount of excess Mn is also incorporated as an impurity in the SLs and the QLs in that sample. Figure 2b shows the STEM image for another sample that contains 71% SLs. Interestingly, that sample was grown with a similar Mn BEP ratio as the sample in Figure 2a (0.10) and has similar total Mn content of 27.4% as obtained from EDX. The lower number of SLs in this sample suggests that an even larger fraction of the Mn was incorporated as impurity atoms (Mn in Sb and Te sites), and specifically, that the QLs may also contain large quantities of substitutional Mn. Figure 2c shows the STEM image for a third sample, which contains only 13% SLs and has a Mn content of 8.1% (from EDX). As for the previous two samples, based on the Mn content from EDX, a larger amount of SLs (over 50%) would be expected if all the Mn atoms were incorporated to form stoichiometric SLs. The presence of excess Mn throughout some regions of the sample can be seen in Figures 2d and 2e, which show the Mn EDX cross sectional scan overlaid on the STEM image in two different regions of the sample of Figure 2c. In the region depicted in Figure 2d the Mn is mostly confined to the SLs, whereas the Mn atoms in Figure 2e are more randomly distributed throughout the crystal lattice and occupy other sites, while also forming SLs.

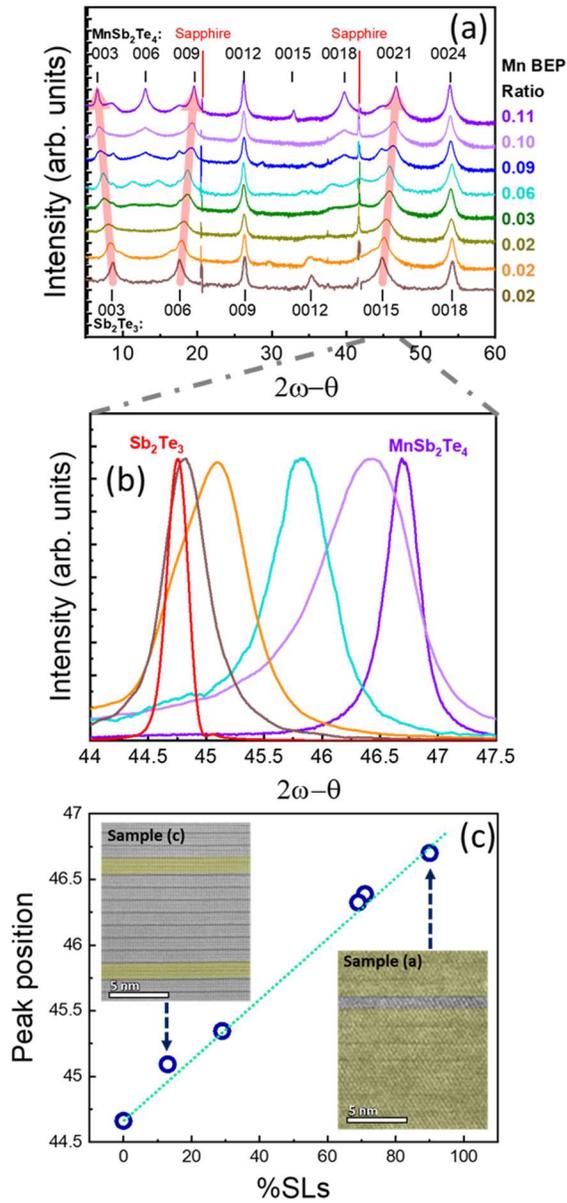

Figure 3: (a) HR-XRD measurements of samples grown with increased Mn BEP Flux ratios. Peak positions for $Sb_2Te_3$ and $MnSb_2Te_4$[13] are noted as black lines below and above the measurements respectively. Mn BEP ratios for each sample are given in the same color as the sample scan to the right of the plots. (b) HR-XRD measurements of the region around the 0015 ($Sb_2Te_3$) to 0021 ($MnSb_2Te_4$) peaks of samples. The peak positions shift away from the position

of $Sb_2Te_3$ as the Mn level increases, indicating the change in the crystal structure. (c) Peak position of the samples plotted as a function of their calculated %SLs from STEM (blue). A linear fit line is plotted forced to go through 0% SLs. Insets are the STEM images for the corresponding samples shown in Figures 2a and 2c.

All of the samples were characterized by high resolution X-ray diffraction (HR-XRD). A selection of the HR-XRD scans for samples ranging in Mn BEP fraction from 0.02 to 0.11 are shown in Figure 3a, where the sample presented in Figure 2a is the one corresponding to the top XRD scan. The expected position of the peaks for pure $Sb_2Te_3$ (bottom) and pure $MnSb_2Te_4$ (top), are also indicated in the figure. The XRD scans are ordered by the Mn content of the samples, from the lowest Mn content at the bottom to the highest at the top. The Mn BEP ratio used during growth is given on the right of each scan. It is evident that as the content of the Mn increases, there is a gradual change in the HR-XRD from that of a QL-based $Sb_2Te_3$ crystal structure to a $MnSb_2Te_4$ (SL-based) structure. The $Sb_2Te_3$ peaks corresponding to the (003), (006) and (0015) planes show a clear shift as the Mn fraction increases, evolving into the (003), (009) and (0021) peaks of $MnSb_2Te_4$, respectively, as can be seen by the red arrows superimposed on the plots. At intermediate Mn flux fractions (0.03-0.06), two weak peaks appear at around $2\theta-\omega$ ~10 - 15°, as previously reported by others[17]. Those authors proposed that this feature indicates a change of structure from $Sb_2Te_3$ to close to $MnBi_4Te_7$. At the high Mn fraction levels (0.09-0.11), a single peak that corresponds closely to $MnSb_2Te_4$ (006) and (0018) planes appears. To further illustrate the gradual shift in the XRD peaks, Figure 3b shows an expanded view of the region around the (0015) plane of $Sb_2Te_3$ for a few selected samples. A clear trend is visible in the shift of the peak position as the Mn content increases, ranging from the $Sb_2Te_3$ (0015) peak position to the $MnSb_2Te_4$ (0021) peak position.

Figure 3c shows the relationship between the shift of the HR XRD (0015) peak and the %SL calculated from the STEM images. The calculated %SLs in the structure measured from the STEM images plotted as a function of the peak position shows a very good linear correlation. Thus, the shift in the XRD peak position is a direct measure of the %SL in the structure, and the linear relationship obtained, as given by Equation (1), can be used to estimate the %SL in all of our samples based on the HR-XRD (0015) peak position.

$$(1) \quad \%SL = \frac{(XRD\ peak\ position) - 44.68}{0.023}$$

While this shift correlates with the increased presence of SLs in the samples, it is not an indication of the total Mn content. This is exemplified by the results for the sample shown in Figure 2a, which although having nearly the same peak position as that reported for pure (stoichiometric) $MnSb_2Te_4$, its Mn concentration according to EDX is almost twice as high as that of stoichiometric $MnSb_2Te_4$ (14.3 % Mn). Thus, we conclude that our samples, grown under the conditions described above, are Mn-rich and likely contain a fair amount of disorder in the form of Mn impurity concentration.

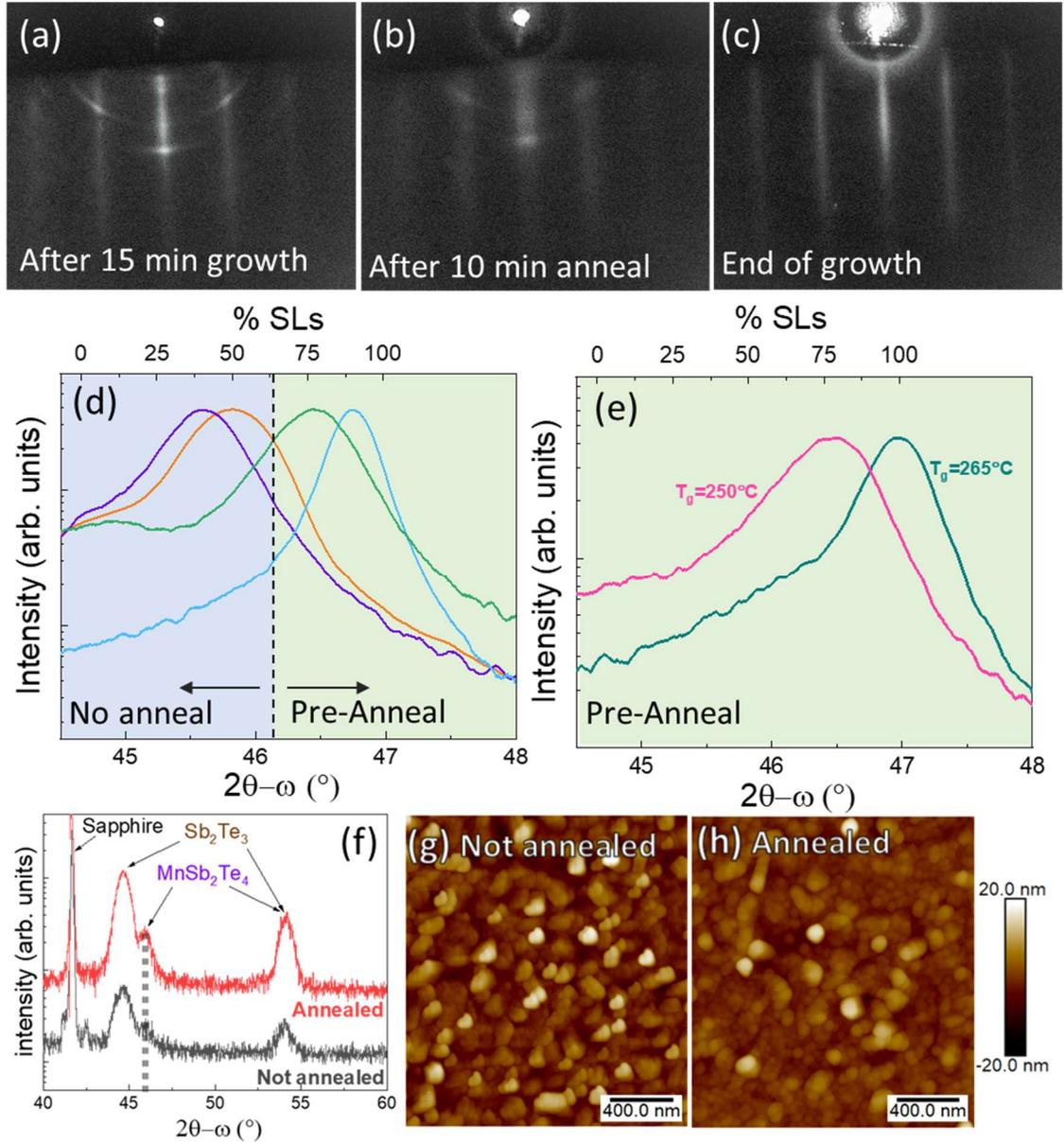

Figure 4: (a-c) RHEED images of a sample surface before (a), and after (b), pre-annealing and at the completion of the growth (c). (d) HR-XRD measurements of the 0015 (0021) peak of samples grown with and without the pre-anneal step. The blue and green shading distinguishes between scans for samples with the pre-anneal step (green) from those without (blue). (e) HR-XRD measurements of the 0015 (0021) peak of two samples grown with standard ($T_g$= 250°C) and increased ($T_g$= 265°C) growth temperatures. The green shading indicates that both samples

in (e) were grown using the pre-anneal step. (f) HR-XRD scans of two thin layers grown for 15 minutes with (red) and without (black) an annealing step. Peak position attributed to $MnSb_2Te_4$ (0021) is marked with dotted line. (g,h) 2μm x 2μm AFM images of the samples in (f). Both AFM images are based on the given scale bar.

In an attempt to reduce the amount of apparent disorder in the crystals, a modified growth procedure was developed. Two changes were implemented. First, an annealing step was added after an initial thin portion of the layer was grown (from here on referred to as *pre-anneal step*). Annealing has been used often in MBE and other growth techniques to improve crystalline quality[18-19]. Using an anneal step after a short growth to promote SLs formation has been recently reported by others[20]. In our experiments, the sample growth began the same way as before, with the thin $Sb_2Te_3$ LTB followed by an interruption of the growth and the initiation of growth of the Mn-containing layer at the desired growth temperature ($T_g$). After 15 minutes, the growth is again stopped, and the sample temperature is raised to 300°C for 10 minutes under a Te flux (pre-anneal step). The temperature is then lowered back to $T_g$ and growth of the Mn containing layer is resumed for the desired total growth time.

As a first assessment of the effect of the anneal step, we examined the evolution of the RHEED pattern. The RHEED image of the sample after 15 minutes of growth, but before the pre-annealing step (Figure 4a), often shows noticeable rings, suggesting some polycrystalline growth. After completing the 10 min. annealing step, the RHEED pattern typically improves (Figure 4b), showing less evidence of polycrystalline growth. Finally, at the end of the full layer growth, the RHEED improves significantly (Figure 4c). More importantly, implementation of the pre-annealing step had other effects on the sample that is grown over the pre-annealed thin layer.

Figure 4d shows a comparison of the (0015) peak of the XRD scans of four samples, two grown with the pre-anneal step and two grown without. All samples in Figure 4d were grown using a similar Mn flux fraction, $T_g$ and Te overpressure. The figure shows a clear shift toward larger angle in the (0015) XRD peaks for the two samples grown with the pre-anneal step, suggesting that larger %SLs form in the samples grown over the pre-annealed thin layer. The calculated %SLs for each sample of 4d, summarized in Table 1 as samples 1 through 4, increase from 40-50 %SLs to 70-90 %SLs with the pre-annealed layer (samples 3 and 4), indicating that more of the Mn is incorporated in the form of SLs when the layer growth occurs over a thin pre-annealed layer, consequently reducing the incorporation of substitutional Mn into QL and SL.

Table 1: Calculated %SLs for samples grown under different growth conditions.

| Sample | Pre-annealed Layer | Mn BEP ratio | $T_g$ (C) | % SLs |
|---|---|---|---|---|
| Effect of pre-annealed layer (Figure 4a) | | | | |
| 1 | No | 0.060 | 255 | 51 |
| 2 | No | 0.060 | 260 | 41 |
| 3 | Yes | 0.060 | 255 | 72 |
| 4 | Yes | 0.065 | 255 | 92 |
| Effect of increased growth temperature (Figure 4b) | | | | |
| 5 | Yes | 0.090 | 250 | 79 |
| 6 | Yes | 0.090 | 265 | 100 |

In some samples, in addition to the pre-anneal step, the $T_g$ was also increased to 265-270°C. The (0015) peak of the HR-XRD scans of two samples grown with a pre-anneal step and the same value of Mn BEP, but with two different $T_g$ values of 250°C and 265°C, are presented in Figure 4e. A further increase in the %SLs was observed for the sample grown at the higher growth temperature of 265°C indicated by the shift of the (0015) peak. The %SLs for these two samples, calculated from the XRD peak position, are given Table 1, as samples 5 and 6. Sample 6, grown at 265°C, shows a further increase of the %SLs bringing its composition up to 100% SLs.

In order to understand the effect of the pre-anneal step, an experiment was performed in which a comparison was made between a thin layer grown for 15 minutes and annealed for 10 minutes at 300°C, and another thin layer that was also grown for 15 minutes but not annealed. The HR-XRD of the two samples are shown in Figure 4f. The stronger and sharper XRD peaks for the annealed sample suggest improved crystallinity. However, the positions of the (0015) peak for the two samples, identified by a dashed line in the figure, are the same, indicating the same %SL for both samples. The AFM images of the surfaces of the two samples (Figures 4g-h) show that the surface of the annealed sample (Figure 4h) is smoother (rms = 3.0nm) than that of the sample (Figure 4g) that was not annealed (rms = 4.2nm). The data of Figures 4f-h suggests that the pre-anneal step improves the crystalline quality of the annealed thin layer and makes its surface smoother, but it does not change the %SL of the thin annealed layer. Thus, we conclude that the increased %SLs obtained in the layer grown over the pre-annealed thin layer must be explained by a different mechanism not evident from our experiment.

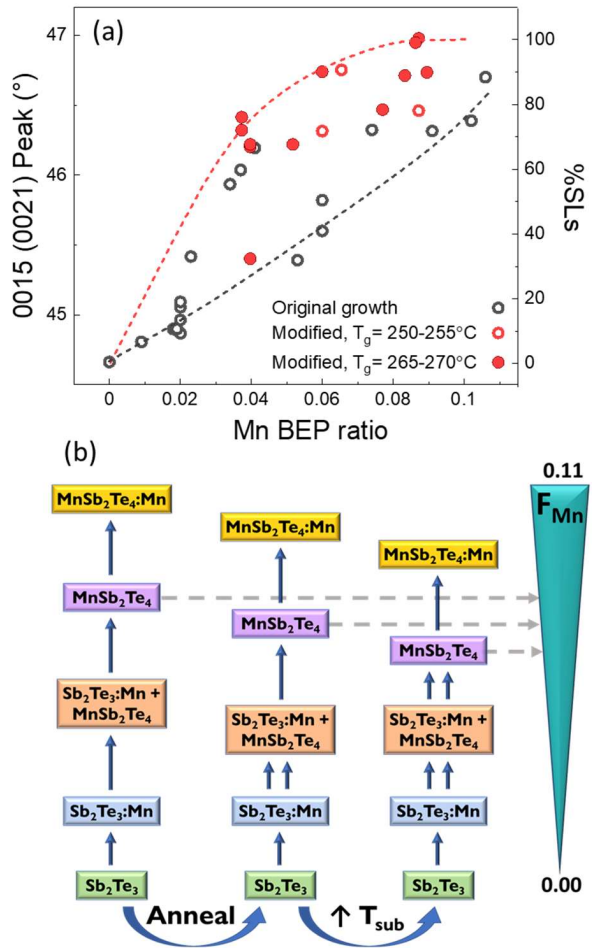

Figure 5: (a) Peak position of all samples (and %SLs), as a function of the Mn BEP ratio. Samples grown under the original growth conditions are shown in black, while samples grown with the modified conditions are given in red: samples with the pre-annealed layer only (red open circles) and those with pre-annealed layer and increased growth temperatures (red full circles). Black and red dashed lines represent the high and low boundaries of that data, highlighting the generally higher %SL obtained with the modified conditions. (b) Illustration of the proposed stepwise growth process for the material formed as a function of increasing Mn BEP ratios (green cone), and the changes that the two modification steps used: pre-annealing step and increased growth temperatures, produce on the growth. The grey arrows indicate the lower Mn flux needed to achieve stoichiometric MnSb2Te4 when the growth modifications steps are introduced.

The relationship between the Mn BEP fraction and the %SL of all the samples grown is illustrated in Figure 5a, which shows the samples grown using the original growth conditions as black open circles, the samples grown using the pre-anneal step as red open circles and those grown using the pre-anneal step as well as a higher $T_g$ as full red circles. Within the range of MBE growth temperatures used here (250-270°C), it can be assumed that all the Mn incorporates in the layer (i.e., Mn will not re-evaporate from the growth surface). For the samples grown under the original growth conditions (black circles), there is a reasonable correlation between the Mn BEP ratio and the XRD peak shift (or %SLs), but a large amount of scatter is present in the data. The scatter is consistent with our assessment that Mn is incorporated into the crystal in different ways, and that a significant fraction of Mn is being incorporated as substitutional dopants without modifying the crystal structure (i.e., without forming SLs).

We also observe that the samples grown with the modified growth conditions shown by the points in red, all have higher %SLs for the same Mn BEP ratio. This is consistent with our conclusion that the pre-anneal step promotes the formation of SLs in the subsequent layer, rather than its incorporation in the lattice as a dopant. Within the red data points, the samples grown at a higher $T_g$, illustrated by the filled red circles, typically result in even higher %SLs for the same Mn BEP during growth.

Our observations are summarized schematically in Figure 5b. The Mn BEP ratio used during growth is illustrated by the green cone on the right, increasing from values of 0.00 to 0.11 in our growths. At very low Mn BEP ratios, only $Sb_2Te_3$ doped with Mn ($Sb_2Te_3$:Mn) can form. As more Mn is used during growth, (as the Mn BEP ratio increases), SLs can be formed resulting in mixed QL and SL structures. The implementation of a pre-annealing step results in SLs formed at lower Mn BEP ratio. Based on these observations we propose the following tentative mechanism. We assume that a critical quantity of excess Mn atoms must be present at the surface of the sample in order to form a SL rather than incorporate as impurities in the QLs. During growth, the Mn incorporates into the QL as impurity atoms and some excess Mn accumulates on the growing surface until there is a sufficient quantity required to form a SL. We propose that the pre-annealing step facilitates the accumulation of Mn on the surface, thus favoring the formation of SLs at lower Mn BEP ratios as compared to when no pre-annealing step is performed. Once this excess Mn is produced, it is more likely for an excess of Mn to remain on the surface as the layers grow, promoting the formation of SLs throughout the growth.

The effect of increasing the growth temperature is also illustrated in Figure 5b. An increase in the %SL formation for the same Mn BEP fraction is also observed by the use of a higher $T_g$. The reason for this enhancement might also be related to the ease of Mn accumulation on the surface at a higher temperature, since at a higher temperature Sb may desorb from the surface leading to higher Mn concentration. However, the growth temperature effect may also be due to the higher formation energy of MnSb$_2$Te$_4$[21], which may be favored at higher T$_g$.

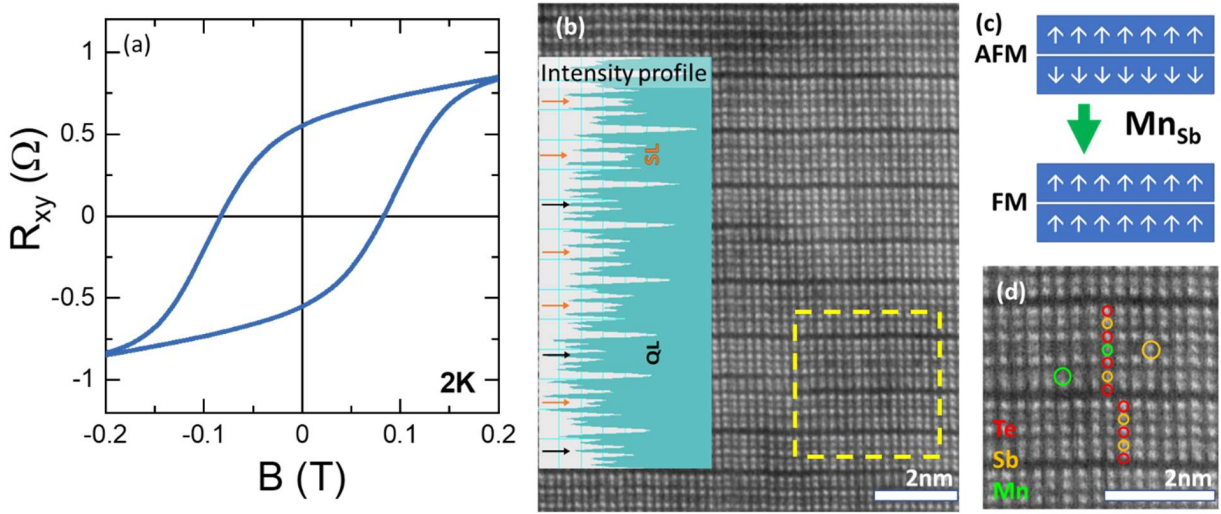

Figure 6: Analysis of a sample with 72% SLs; (a) Hysteresis loop of the Hall resistance Rxy as a function of the magnetic field, B, measured at 2K. (b) HAADF STEM image of the sample. Inset shows an intensity profile aligned with the image. (c) Illustration of proposed mechanism for ferromagnetic alignment of MnSb2Te4 SL produced by the formation of MnSb antisites. (d) Enlarged image of the region marked by the yellow square in (b). The atomic arrangements of SLs and a QLs are illustrated by small red (Te), orange (Sb) and green (Mn) circles. Larger orange and green stand-alone circles suggest Sb at a Mn site, and Mn at an Sb site, respectively, based on the respective image intensity.

Magnetotransport measurements on our samples revealed that all samples having at least a few %SLs exhibited ferromagnetic behavior, as demonstrated by the presence of a hysteresis loop near ~0T magnetic field (B). This result is counter to the expected anti-ferromagnetic behavior predicted by DFT calculations for pure $MnSb_2Te_4$[22]. A typical Hall resistance as a function of magnetic field for a sample that has 72% SLs is shown in Figure 6a. The STEM image, in Figure 6b, confirms the %SLs, and shows no significant defects. The Mn content in this sample is ~21% (as measured by EDX), which is about 50% higher than the expected value for ideal (stoichiometric) $MnSb_2Te_4$ crystal with 100% SLs. This indicates that, as found previously in the samples of Figure 2, there is significant excess Mn in the crystal lattice. Ferromagnetism in Mn-rich $MnSb_2Te_4$ has been recently reported and is attributed to Mn-Sb site exchange (illustrated in Figure 6c)[10].

In order to determine the location of the excess Mn in the crystal in our samples, intensity profile measurements of the HAADF images, which are sensitive to atomic number, were performed. The result, depicted in the inset of Figure 6b, shows a combination of SLs evident by a sequence of seven intensity peaks, one for each atomic plane, and QLs, evident by the presence of five peaks corresponding to the five atomic planes. The profile is aligned with the section of the STEM image where the profile data was collected for ease of interpretation. The center peak of each SL or QL unit is marked by orange or black arrows, respectively. The SLs have a drop in intensity in the peak corresponding to the middle layer due to the small mass of the Mn atoms, suggesting predominantly Mn in that layer, as expected. However, the peaks corresponding to Sb and Te in the SLs show some variations in intensity, which possibly can be attributed to Mn site exchange. By contrast, the QLs show uniformly high intensity for all the peaks, as expected from the closely similar atomic mass of Te and Sb. Further observation of intensity variations in the STEM image also suggests Mn substitution in some Sb sites and Sb substitution in some Mn sites of the SL, examples of which are illustrated by the large orange and green circles in the expanded region of the STEM image shown in Figure 6d. Thus, we conclude that the ferromagnetic behavior of our MnSb$_2$Te$_4$ could also be due to Mn-Sb site exchange. A detailed study of the magnetic properties of our samples is currently underway and is beyond the scope of this paper.

Conclusions:

In this work, we have investigated the growth by MBE of (Sb$_2$Te$_3$)$_x$(MnSb$_2$Te$_4$)$_y$ structures on sapphire substrates. Increasing the flux fraction of Mn (or Mn BEP fraction) during growth results in a gradual change of the composition of the structure from all Sb$_2$Te$_3$ QLs (y = 0) to all SLs (x = 0). The presence of significantly more Mn in the samples than what is accounted for by the percent (%) of SLs, suggests that Mn also incorporates as a dopant forming Sb$_2$Te$_3$:Mn and MnSb$_2$Te$_4$:Mn. We conclude that under our initial growth conditions, controlling the Mn flux ratio is not sufficient to accurately determine the resulting (Sb$_2$Te$_3$)$_x$(MnSb$_2$Te$_4$)$_y$ structure, and that a large degree of excess Mn is incorporated in the layers. We also analyzed the evolution of the XRD scans of the layers as the Mn content is increased. The results show a gradual shift of several peaks from the Sb$_2$Te$_3$ position to the MnSb$_2$Te$_4$ position. We show that this shift can be used to calculate the %SLs in the resulting crystal.

By modifying the growth conditions with the incorporation of a pre-annealed thin layer at the beginning of growth, a significant increase of %SL formation for the same Mn flux fraction used during growth was observed in the overall layer. Thus, with the pre-annealed layer, the incorporation of Mn as a structural element to form SLs, is favored. We propose a mechanism consistent with our observations based on the assumption that a local excess of Mn is needed at the growth front for the formation of SLs during growth. We suggest that the formation of a Mn-rich growth surface is facilitated by the pre-annealing step. A further increase in the %SLs for the same Mn flux fraction was observed by also increasing the growth temperature.

Hall resistance as a function of magnetic field measurements of our samples show that the samples are ferromagnetic, which may be in part related to the excess Mn incorporation in our layers. We present an HAADF intensity profile measurement to show evidence of substitution of Mn in Sb sites, as well as Sb in Mn sites. Mn/Sb anti-sites have been reported to lead to ferromagnetic behavior. The demonstrated control of the composition of Mn containing $Sb_2Te_3$ structures and the evidence for reduced excess Mn in samples grown under the modified growth conditions explored provide essential insights for the controlled growth by MBE of these highly promising and technologically relevant materials.

**Acknowledgements**

This work was supported by NSF Grant Nos. DMR-1420634 (MRSEC PAS[3]), DMR-2011738 (PAQM), HRD-1547830 (CREST IDEALS) and HRD-2112550 (Phase II CREST IDEALS). The authors would like to acknowledge the Nanofabrication Facility of the CUNY Advanced Science Research Center (ASRC) for instrument use, scientific and technical assistance.

*mtamargo@ccny.cuny.edu